\documentclass[iicol]{sn-jnl}

\usepackage{graphicx}%
\usepackage{multirow}%
\usepackage{amsmath,amssymb,amsfonts}%
\usepackage{amsthm}%
\usepackage{mathrsfs}%
\usepackage[title]{appendix}%
\usepackage{xcolor}%
\usepackage{textcomp}%
\usepackage{manyfoot}%
\usepackage{booktabs}%
\usepackage{algorithm}%
\usepackage{algorithmicx}%
\usepackage{algpseudocode}%
\usepackage[title]{appendix}%
\usepackage{listings}%
\usepackage{comment}
\usepackage{natbib}
\usepackage{censor}
\usepackage{subcaption}

\begin{document}

\title[Article Title]{People Perceive More Phantom Costs From Autonomous Agents When They Make Unreasonably Generous Offers}


\author*[1]{\fnm{Benjamin} \sur{Lebrun}}\email{benjamin.lebrun@pg.canterbury.ac.nz}

\author[2]{\fnm{Christoph} \sur{Bartneck}}\email{christoph.bartneck@canterbury.ac.nz}

\author[3]{\fnm{David} \sur{Kaber}}\email{david.kaber@oregonstate.edu}


\author[1]{\fnm{Andrew} \sur{Vonasch}}\email{andrew.vonasch@canterbury.ac.nz}

\affil[1]{\orgdiv{School of Psychology, Speech and Hearing}, 
\orgname{University of Canterbury},
\city{Christchurch},
\country{New Zealand}}

\affil[2]{\orgdiv{Department of Computer Science and Software Engineering},
\orgname{University of Canterbury},
\city{Christchurch},
\country{New Zealand}}

\affil[3]{\orgdiv{College of Engineering},
\orgname{Oregon State University},
\city{Corvallis},
\state{Oregon}, \country{USA}}




\abstract{
People often reject offers that are too generous due to the perception of hidden drawbacks referred to as ``phantom costs.''
We hypothesized that this perception and the decision-making vary based on the type of agent making the offer (human vs. robot) and the degree to which the agent is perceived to be autonomous or have the capacity for self-interest.
To test this conjecture, participants (N = 855) engaged in a car-buying simulation where a human or robot sales agent, described as either autonomous or not, offered either a small (5\%) or large (85\%) discount.
Results revealed that the robot was perceived as less self-interested than the human, which reduced the perception of phantom costs. While larger discounts increased phantom costs, they also increased purchase intentions, suggesting that perceived benefits can outweigh phantom costs.
Importantly, phantom costs were not only attributed to the agent participants interacted with, but also to the product and the agent’s manager, highlighting at least three sources of suspicion.
These findings deepen our understanding of to whom people assign responsibility and how perceptions shape both human-human and human-robot interactions, with implications for ethical AI design and marketing strategies.
}




\maketitle

\section{Introduction}
People often suspect ulterior motives when others make unreasonably generous economic offers without a clear rationale \cite{Vonasch2024, HOSEVonasch}. This happens in part because people typically expect others to act in their self-interest \citep{Miller1999}. For instance, people are more likely to reject a cookie from a stranger who offers to pay them to eat it, imagining that something must be wrong with it \citep{Vonasch2024}. This inferred downside is referred to as a ``phantom cost'', which may include ulterior motives and risks such as pranks or poisoning. Such judgments are grounded in people's interpretations of behaviors, which they use to infer others' mental states or intentions in order to explain and predict their behaviors \citep{Dennett1987, Gray2007}.

Interestingly, this perception extends beyond human interactions. A prior study showed similar perceptions of phantom costs when the offer came from a robot---an agent that arguably lacks a mind \citep{Gray2007, LeeMindSales2024} \citep{Lebrun2024arXiv}. However, the underlying mechanism remains unclear. According to \cite{Vonasch2024} and their Heuristic of Sufficient Explanation (HOSE) model, phantom costs occur when the agent is overly generous without a clear rationale, while this same agent would have been expected to act selfishly. When a robot makes such offer, the same heuristic might apply, albeit in different ways than with humans. Specifically, individuals may perceive different sources of phantom costs depending on who they think is the actual decision-maker who has ulterior motives: the robot or its operator.

\subsection{Phantom Costs}
Phantom costs can lead individuals to reject economically beneficial offers. \cite{Vonasch2024} demonstrated this through the ``money backfire effect.'' This effect occurs when unreasonably generous incentives decrease the chances of someone accepting an offer, rather than increasing the chances. For example, participants were more likely to accept a cookie when offered alone than when money was added to the deal, for which they were suspicious of. In another scenario, participants rejected a \$15 airline ticket in favor of a \$205 ticket, due to the perception of phantom costs in the former case (such as frequent crashes) \cite{Vonasch2024}.

\cite{Vonasch2024} theorized this phenomenon in their Heuristic of Sufficient Explanation (HOSE) model. Specifically the model posits that when an agent's action seems to lack sufficient explanation, observers seek justifications that frequently are phantom costs, such as the person is not genuinely generous and has ulterior motives that could inflict harms for the recipient. This negatively impacts the decision to accept an offer. However, when a reasonable explanation is provided---e.g., the cheap airline ticket was advertised as having uncomfortable seats---phantom costs are less likely to be perceived, increasing the acceptance rate \citep{Vonasch2024}.
Unselfishness could be another implicit justification for a generous offer, although unlikely in human-human interactions between strangers. Robots, however, due to their programmed nature, are often perceived as lacking ``personal'' goals or interests \cite{Xu2024}, and more helpful and trustworthy \cite{LeeMindSales2024}.

\cite{Lebrun2024arXiv} replicated the cookie paradigm with both a human and a robot. Both produced a money backfire effect. However, the authors mainly focused on the phantom costs related with the offer, not whether the agents were perceived to have ulterior motives. Because robots have less agency than humans \citep{Gray2007}, we believe that robots are perceived as less self-interested, as their actions serve their designers or operators rather than themselves. This perceived lack of self-interest might suffice to justify their generous behavior, reducing perceptions of phantom costs.

\subsection{Self-Interest}
People typically behave to attain personal gains \citep{Cropanzano2005, Dawkins2006, Knoch2006, Shalvi2012}, and expect others to do the same \citep{Miller1999}.
Violation of the expectation of self-interest, especially without a clear rationale, motivates people to seek explanations and perceive phantom costs \citep{Ratner2001, Vonasch2024}.
This expectation of self-interest may change when the agent is a robot. On the one hand, if a robot is perceived as self-interested, then an unreasonably generous offer provided without a clear rationale might potentially invoke a phantom cost or suspicion of the robot.
On the other hand, if the robot is perceived as lacking self-interest, the same offer might not lead to the perception of phantom costs, or at least to a lesser extent than in human interactions. It might be sufficient to believe the robot made the offer because it is meant to serve, not to benefit. 
Yet, \cite{Roselli2023} showed that people attribute intentions to robots somewhat similarly to humans \citep{Marchesi2019, Thellman2017}. More broadly, people often project their own intentions (in context) to other agents \citep{Epley2007}. If so, people should perceive phantom costs similarly in human-human (HHI) and human-robot interaction (HRI). Still, people could perceive phantom costs, not from the robot itself but from the potential human ``behind the scene.''

In the sales industry \cite{LeeMindSales2024}, such as in car sales \cite{Verlegh2004}, self-interest is expected. Salespeople usually earn commissions \cite{Lopez2006}, and large discounts increase suspicion, reducing purchase intentions \cite{Prasetyo2020, Lee2014}.
A robot in the role of car sales agent might also increase suspicion, especially if the dealership aims to maximize profits. Robots selling cars could seem far-fetched for present day applications. However, in Commack, New York, a dealership has employed a sales robot named Promobot to assist customers, collect their information, and contact a salesperson when required \citep{Promobot2020}. 

However, a crucial difference between salespeople and sales robots is the degree of autonomy, which may alter buyer perception of phantom costs. For instance, Promobot must contact a salesperson to make the actual sale. Autonomy requires making choices that reflect one's values and own agency (see \cite{Meyers1989, Thalos1997}). Promobot does not. One might expect Promobot not to act in its own self-interest, but in the interest of its employer: the car dealership. If Promobot makes an unreasonably generous discount, a customer might imagine a phantom cost, not because they think Promobot should be \textit{self}-interested, but because they suspect the robot is programmed to maximize the \textit{dealership's} self-interest.
This means the customer perceives the dealership as the agent making the offer and using the sales agent as a proxy for benefit. For this reason, phantom costs may also be directed toward the dealership's manager.

\subsection{Preliminary Study}
We conducted a preliminary study to test whether people perceive fewer phantom costs when interacting with a robot and whether this increased the likelihood of accepting unreasonably generous offers. This study was motivated by the idea that robots, unlike humans, are not typically perceived as self-interested. Hence, an unreasonably generous offer made by a robot should elicit fewer phantom costs than the same offer made by a human. Initial results supported this. So, we conducted a full experiment to explore how the type of agent and degree of agent autonomy might influence the perception of phantom costs. 

\subsection{The Present Research}
The present research builds on \cite{Lebrun2024arXiv} and aims to further identify psychological mechanisms underlying the perception of phantom costs in both HHI and HRI. We conducted this study using vignette-based scenarios: a human or a robot selling cars at a dealership for either a reasonable or unreasonably generous price. The latter condition was expected to elicit phantom costs due to a lack of sufficient rationale for the pricing.

We hypothesized that the perception of phantom costs may differ between HHI and HRI. Since robots are non-living agents with programmed behaviors, phantom costs in HRI may not be (only) attributed to the robot itself but to a presumed human programmer/operator. We also investigated agent autonomy: the agent was either described as making decisions or following a manager's rules. Our manipulation of autonomy followed \cite{Kaber2017}'s framework which defines autonomy in terms of three facets: self-governance (i.e., absence of external control and the use of cognitive capabilities to perform a task, such as learning, initiating tasks and using strategies), viability (i.e., having the required characteristics to function in a domain), and independence (i.e., agent's capacity to perform its task).

We also explored the relationship between phantom costs and trust, as an additional response to HHI and HRI. Trust is defined as the confidence that an agent will act to support an individual’s objectives in situations where there is uncertainty and potential risk \citep{LeeSee2004}. Our prior work \cite{Lebrun2024arXiv, Lebrun2025Replication} indicated that trust mediates human decision making in interaction with other agents when there are perceived phantom costs. In specific, \cite{Lebrun2025Replication}  showed that trust mediated decisions of whether to accept economic offers.
This aligns with research exploring how explanations---that the HOSE model requires to be sufficient---can foster trust \citep{Edmonds2019, Sanneman2020, Setchi2020} and guide reasoning \citep{Lombrozo2006}.

By addressing limitations of prior work, the current study aims to clarify whether robots elicit fewer phantom costs due to being perceived as less intentional or self-interested.
The present study provides further insights into how phantom costs apply to HRI, how they differ from or resemble those in HHI, and what are the sources of phantom cots that have to be taken into account during an interaction: the product, the agent, or their manager. The implications of this study are twofold. First, results may assist companies in improving robot marketing strategies. Second, the study provides guidance to robot and AI designers on how such agents could behave in explaining behaviors to promote likelihood of human trusted engagement.
Hypotheses and research questions are shown in \autoref{tab:hypothese}.

\begin{table*}[h!]
\caption{Summary of hypotheses (H) and exploratory research questions (RQ)}
\label{tab:hypothese}
\centering
\resizebox{\textwidth}{!}{
\begin{tabular}{cll}
\toprule
\textbf{\#} & \textbf{Topic} & \textbf{Hypothesis/Research Question} \\ \midrule
H1 & Manipulation check: Self-Interest & \parbox[t]{10cm}{Humans are perceived as more self-interested than robots.} \\ \midrule
H2 & Manipulation check: Perceived Autonomy & \parbox[t]{10cm}{Autonomous agents are perceived as more autonomous than non-autonomous agents.} \\ \midrule
H3 & Phantom costs toward the \textit{Agent} & \parbox[t]{10cm}{Higher when: \\
\begin{enumerate}[(a)]
    \item Agent is a \textbf{human} rather than a robot.
    \item Agent offers a \textbf{large} discount without sufficient explanation.
    \item Agent is \textbf{autonomous} rather than non-autonomous.
\end{enumerate}} \\ \midrule
H4 & Phantom costs toward the \textit{Product} & \parbox[t]{10cm}{Higher when the product is sold with a \textbf{large discount} rather than a small one.} \\ \midrule
H5 & Phantom costs toward \textit{Agent's Manager} & \parbox[t]{10cm}{ \begin{enumerate}
    \item[(a)] \textbf{Do not differ} between a human and robot agent. \vspace{2mm}
\end{enumerate}Higher when: \\
\begin{enumerate}
    \item[(b)] A \textbf{large discount} is offered.
    \item[(c)] Agent is \textbf{non-autonomous} rather than autonomous.
\end{enumerate}} \\ \midrule
H6 & Mediation by Self-Interest & \parbox[t]{10cm}{The \textbf{perception of self-interest} mediates phantom costs perception.} \\ \midrule
H7 & Mediation by Phantom Costs & \parbox[t]{10cm}{\textbf{Phantom costs perception} mediates \textbf{participants' purchase intentions}.} \\ \midrule
H8 & Purchase Intention & \parbox[t]{10cm}{Higher when: \\
\begin{enumerate}[(a)]
    \item Agent is a \textbf{robot} rather than a human
    \item Agent offers a \textbf{small} rather than a large discount
    \item Agent is \textbf{autonomous} rather than non-autonomous
\end{enumerate}} \\ \midrule
H9\footnotemark[1] & Phantom costs and Trust & \parbox[t]{10cm}{There is a relationship between \textbf{phantom cost perception} and \textbf{trust}.} \\ \midrule
RQ1\footnotemark[1] & Mediation of Autonomy through Self-Interest & \parbox[t]{10cm}{Does \textbf{perception of self-interest} mediate the relationship between \textbf{Autonomy} and \textbf{Perceived Autonomy}?} \\ \midrule
RQ2\footnotemark[1] & Influence of Income & \parbox[t]{10cm}{Does \textbf{participants' income} influence their decisions to buy a car?} \\ \bottomrule
\end{tabular}
}
\footnotesize{\textit{Notes.} \footnotemark[1]: Data analyses for the exploratory hypotheses and research questions are shown in the Appendices.}
\end{table*}

\section{Method}
We conducted a 2 (Agent: Human vs Robot) $\times$ 2 (Autonomy: Non-autonomous vs Autonomous) $\times$ 2 (Discount: Small vs Large) between-participants design.
The pre-registration is available at \url{https://aspredicted.org/27J_91B}. All data and data analyses are available at \censor{OSF link xxx}.

\subsection{Material}
\subsubsection{Scenario}
We presented participants with text and image-based vignettes to characterize the type of agent and autonomy as shown in \autoref{tab:vignette}. The vignettes were presented at the top of an online survey page so that participants could read it and return to the content as needed. The initial price of the car identified in the vignette (\autoref{fig:Conditions_pictures}) was determined by examining similar-looking cars, as listed by American dealerships. To minimize any framing effect \citep{Kahneman1979}, prices were described using percentages \citep{Morwitz1998, Russo1990} vs. absolute values. We maintained consistent discount ratios of $\sim$5\% and ~85\% for the small and large discounts, respectively, as presented in the vignettes.


\begin{table*}[htbp]
\caption{Vignettes representing the two levels of autonomy}
\small
\renewcommand{\arraystretch}{1.2}
\begin{tabular}{|p{0.47\textwidth}|p{0.47\textwidth}|}
\hline
\textbf{Non-autonomous agents} & \textbf{Autonomous agents} \\
\hline
\multicolumn{2}{|p{0.94\textwidth}|}{You are thinking about buying a used car from a dealership.} \\
\multicolumn{2}{|p{0.94\textwidth}|}{As you enter the dealership, a sales [person/robot] named [Tim/Salesbot] introduces [himself/itself].} \\
\hline
[Tim/Salesbot] is a [person/robot] of limited competence and independent thought. [Tim/Salesbot] has [received training on/programming for] selling cars, but the dealership manager (Steve) keeps a close eye on [Tim/Salesbot] in negotiating with customers in case [he/it] makes a mistake. Although [Tim/Salesbot] can communicate with customers, [Tim/Salesbot] needs to negotiate with Steve on the price for the customer.
&
[Tim/Salesbot] is a highly competent [person/robot] capable of thinking [for himself/on its own], without needing any input or guidance from the dealership manager (Steve) on how to sell cars. Moreover, [Tim/Salesbot] has sold more cars than any other agent at the dealership, and Steve allows [Tim/Salesbot] to set [his/its] own prices without consulting management. \\
\hline
\multicolumn{2}{|p{0.94\textwidth}|}{[Tim/Salesbot] shows you a car that you like, which is in your price range and appears to be in good condition.} \\
\multicolumn{2}{|p{0.94\textwidth}|}{You know the NADA (National Automobile Dealers Association) Blue Book value is typically \$30,000 for the same used car in the same condition.} \\
\multicolumn{2}{|p{0.94\textwidth}|}{[Tim/Salesbot] says:} \\
\hline
``We have it listed at \$31,500, but for you we can make a special deal on it by reducing the price by about [5\%/85\%]. How about \$[29,900/5,000]?'' 
&
``I have it listed at \$31,500, but for you I can make a special deal on it by reducing the price by about [5\%/85\%]. How about \$[29,900/5,000]?'' \\
\hline
\end{tabular}
\footnotesize\textit{Notes.} The agent conditions are presented in the format [Tim/Salesbot] in the vignettes above. The [\$29,900/\$5,000] prices correspond to the small and large discounts, respectively.
\label{tab:vignette}
\end{table*}

\subsubsection{Visual Representation}
\begin{figure}[htbp]
  \subcaptionbox{Robot agent\label{fig:sub1}}{\includegraphics[width=2.5in]{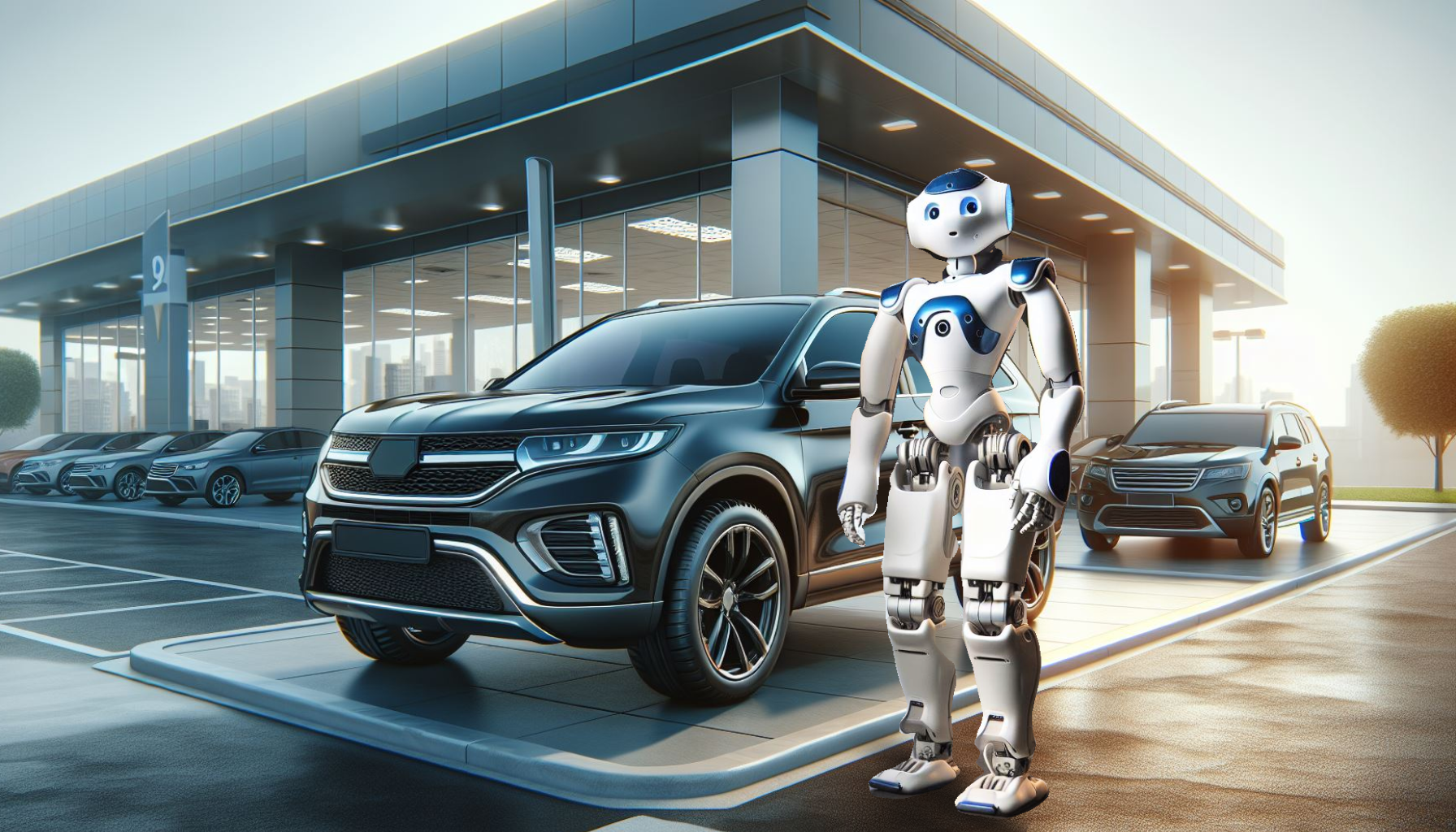}}\hfill%
  \subcaptionbox{Human agent\label{fig:sub2}}{\includegraphics[width=2.5in]{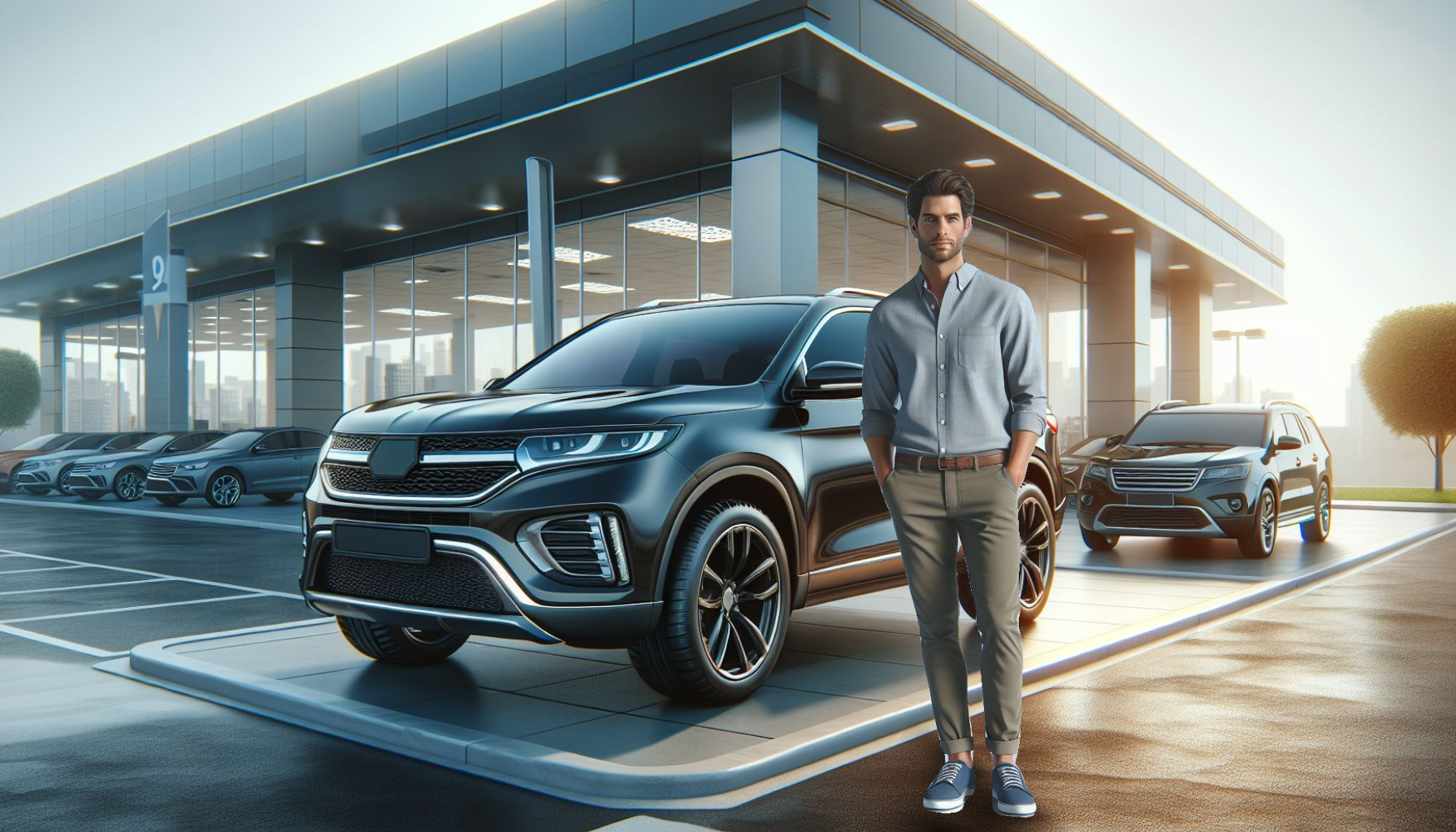}}%

  
  \caption{Pictures Depicting both a Robot (a) and Human (b) Sales Agent at a Car Dealership.}
  \label{fig:Conditions_pictures}
\end{figure}

We used DAlle-E 3 to generate images of a car sales yard. The agents were added to the scene. The pictures represented either a robot (Nao from Aldebaran Robotics, \autoref{fig:sub1}) or a man \autoref{fig:sub2}) in the parking lot of a car dealership. The images were placed below the text of the vignettes.

\subsection{Measurements}
The measurement methods employed in this study are summarized in \autoref{tab:measures}. Note, the 7-point Likert scales ranged from 1 (strongly disagree) to 7 (strongly agree).

\begin{table*}[h!]
\caption{Summary of the Study Measurements, Descriptions, Internal consistency, and Response Formats}
\centering
\resizebox{\textwidth}{!}{
\begin{tabular}{lp{10cm}l}
\toprule
\textbf{Measure} & \textbf{Description} & \textbf{Format} \\ \midrule
Timing & Time on each page; calculated and provided by Qualtrics.& Seconds \\ \midrule
Comprehension Checks & 
First check: Select the correct discounted price (\$5,000, \$7,500, \$10,000, \$29,900, \$31,500).
\newline
Second check: Select the option ``Somewhat disagree.''& Binary: 0 = Fail, 1 = Success \\ \midrule
Decision-making & ``I do not buy the car'' or ``I buy the car.'' & Binary: 0 = No, 1 = Yes \\ \midrule
Phantom Costs & 
Perceptions assessed across three targets:
\begin{itemize}
    \item Car: ``I am confident that the car is of high quality'' (reversed), ``I think it would be risky to buy this car,'' ``I would worry the car might have something wrong with it.''
    \item Sales agent: ``{[}Tim/Salesbot{]} has a hidden agenda in discounting the car.''
    \item Manager: ``The dealership manager has a hidden agenda in discounting the car.''
\end{itemize}
Internal consistency: $\alpha = .860$ (overall), car: $\alpha = .843$. & 7-point Likert scale \\ \midrule
Autonomy & Adapted from \cite{Kaber2017}
\begin{itemize}
    \item Self-governance: ``{[}Tim/Salesbot{]} is able to plan, learn from experience and use strategies to sell the car.''
    \item Independence: ``{[}Tim/Salesbot{]} acts independently to sell the car.''
    \item Viability: ``{[}Tim/Salesbot{]} has the skills to sell the car.''
\end{itemize}
Internal consistency: $\alpha = .830$. & 7-point Likert scale \\ \midrule
Self-interest & Perception of agent's self-interest: ``{[}Tim/Salesbot{]} is motivated by self-interest.'' & 7-point Likert scale \\ \midrule
Trust & 
Adapted from \cite{Lee1992}, repeated for car, agent, and manager:
\begin{itemize}
    \item ``I can count on {[}the car/Salesbot/Tim/the dealership manager{]} to do {[}its/his/their{]} job.''
    \item ``Overall, I trust {[}the quality of the car/Salesbot/Tim/the dealership manager{]}.''
\end{itemize}
Internal consistency: $\alpha = .909$ (overall), car: $\alpha = .941$, agent: $\alpha = .825$, manager: $\alpha = .839$. & 7-point Likert scale \\ \midrule
Demographics & Age, gender, race, and income\footnotemark. & Open-ended / Categorical \\ \bottomrule
\end{tabular}
}
\footnotesize{\textit{Notes.} Income was examined for its potential influence on car-buying decisions.}
\label{tab:measures}
\end{table*}

\subsection{Participants}
Eight hundred eighty-one participants were recruited via \url{https://www.prolific.com} to participate in an online study using the Qualtrics survey tool. Participants were all fluent in English with American nationality.

\subsection{Procedure}
After consenting to participate, participants read the vignette associated with the relevant picture shown in \autoref{fig:Conditions_pictures}. The human agent picture was shown for the salesperson conditions, while the robot agent picture was shown for the sales robot conditions. On the first page of the survey, participants answered the comprehension check and decided whether to buy the car. On the second page, they answered questions regarding their perceptions of phantom costs. They then rated the agent autonomy and self-interest. On a new page, they rated their trust in the car, the agent, and the manager. The second attention check was integrated into the trust questionnaire. Finally, participants provided their demographic information and read a debriefing sheet before being redirected to Prolific to receive their compensation.

\section{Results}
\subsection{Participants}
Twenty-six participants were excluded from the initial sample due to failing the comprehension check. No participants were excluded based on timing. The final sample size included 855 participants: 48.9\% males (n = 418), 49.8\% females (n = 426), and 1.3\% (n = 11) declared their gender to be different. Ages ranged from 18 to 82 ($M = 39.82, SD = 13.28, m = 37$).
Race was distributed as follows:
$\sim$71\% White or Caucasian;
$\sim$14\% Black or African American;
$\sim$9\% Asian;
$\sim$2\% American Indian/Native American or Alaska Native;
$\sim$1\% Native Hawaiian or Other Pacific Islander;
$\sim$2\% selected the option ``Other'';
and $\sim$1\% preferred not to say.
Regarding annual income: $\sim$22\% reported earning less than \$25,000; $\sim$23\% earned \$25,000-\$49,000; $\sim$20\% earned \$50,000-\$74,999; $\sim$13\% earned \$75,000-\$99,999; $\sim$11\% earned \$100,000-\$149,999; $\sim$7\% earned \$150,000 or more; and $\sim$3\% preferred not to say.

\subsection{Items consistency}
Cronbach's alpha values were calculated to assess the internal consistency of each measure. All calculations showed good to excellent reliability (see \autoref{tab:measures}).
To simplify the analyses, we combined the items in each category into a single averaged variable. Unless otherwise specified, all subsequent statistical analyses were based on these computed variables.

\subsection{Perceived Autonomy}
\begin{figure}
    \centering
    \includegraphics[width=\linewidth]{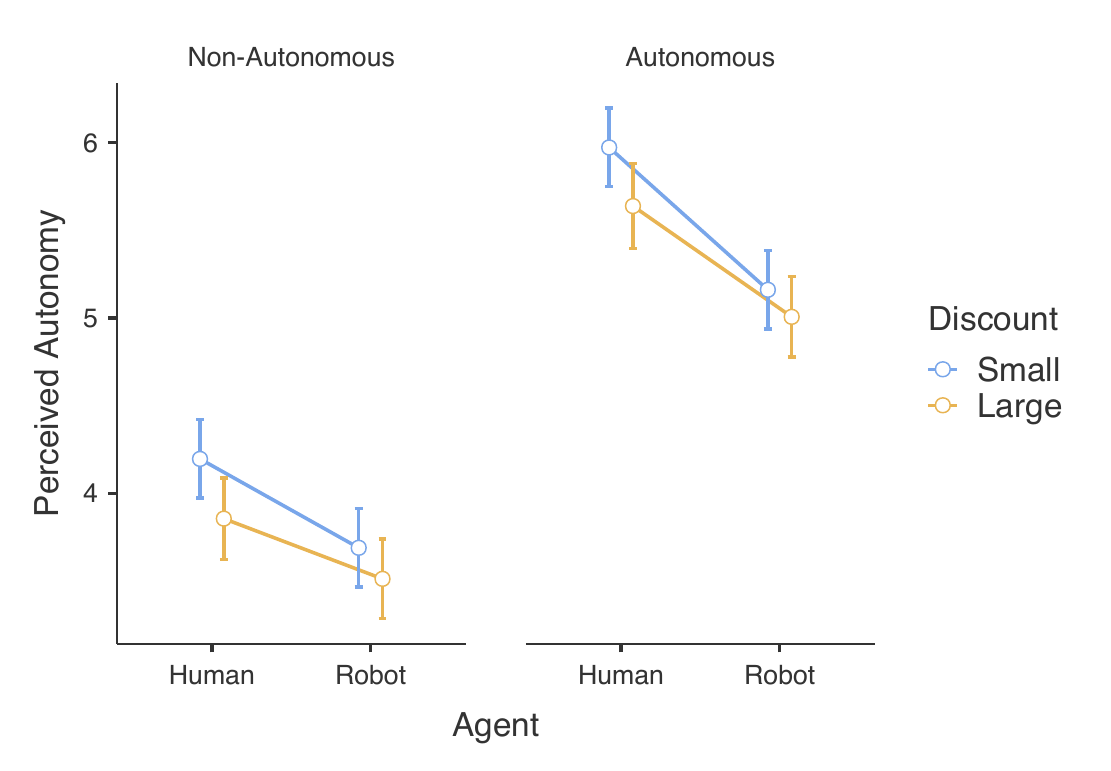}
    \begin{minipage}{0.6\linewidth}
        \centering
        \footnotesize{\textit{Notes.} Error bars correspond to 95\% Confidence Intervals.}
    \end{minipage}
    \caption{Perceived Autonomy as a function of Agent, Discount, and Autonomy.}
    \label{fig:Perceived_autonomy_ANOVA}
\end{figure}

To examine whether we effectively manipulated agent autonomy (independent variable), we applied a three-way ANOVA to perceived agent autonomy as a function of Agent, Discount, Autonomy, and their interactions (see \autoref{fig:Perceived_autonomy_ANOVA}).
Results showed a main effect of Agent ($F(1,847) = 48.59, p < .001, \eta_p^2 = .054$), Discount ($F(1,847) = 9.34, p = .002, \eta_p^2 = .011$), and Autonomy ($F(1,847) = 393.21, p < .001, \eta_p^2 = .317$).
Post-hoc comparisons revealed that the human was perceived as more autonomous ($M = 4.92, SD = 0.06$) than the robot ($M = 4.34, SE = 0.06$), $t(847) = 6.97, p < .001, d = .477$. Agents making small discounts were perceived as more autonomous ($M = 4.75, SE = 0.06$) than those making large discounts ($M = 4.50, SE = 0.06$), $t(847) = 3.06, p = .002, d = .209$.
Agents that we described as autonomous were indeed perceived as autonomous ($M = 5.44, SE = 0.06$), as compared to  the non-autonomous agents ($M = 3.81, SE = 0.06$), $t(847) = 19.83, p < .001, d = 1.358$.

\subsection{Phantom costs perception}
\begin{figure*}[htbp]
  \centering
  \begin{subfigure}[b]{0.49\textwidth}
    \includegraphics[width=\textwidth]{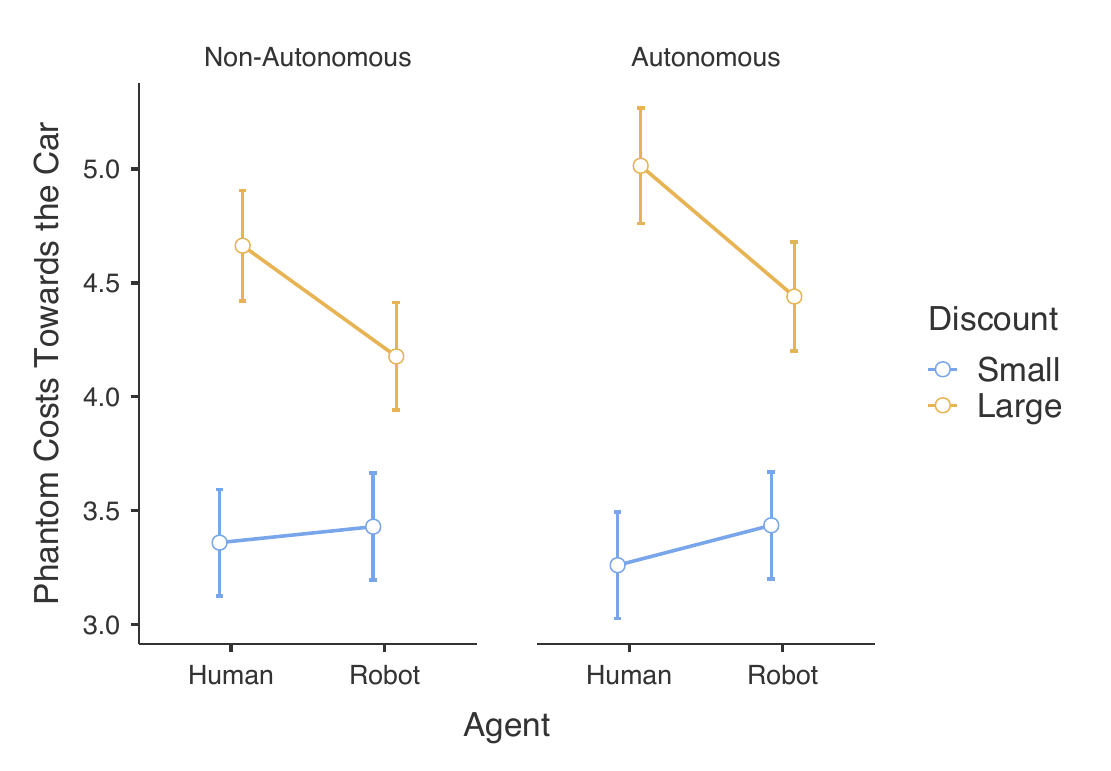}
    \caption{Phantom Costs Towards the Car}
    \label{fig:Phantom_car_ANOVA}
  \end{subfigure}
  \hfill
  \begin{subfigure}[b]{0.49\textwidth}
    \includegraphics[width=\textwidth]{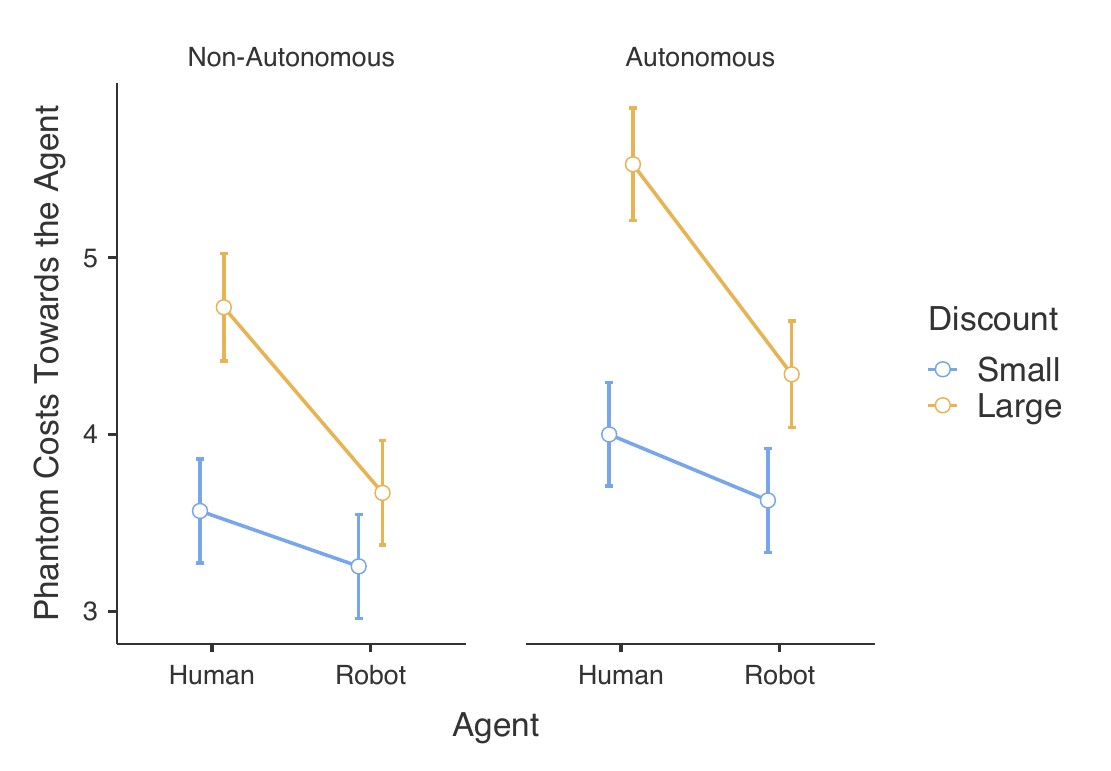}
    \caption{Phantom Costs Towards the Agent}
    \label{fig:Phantom_seller_ANOVA}
  \end{subfigure}

  \begin{subfigure}[b]{0.49\textwidth}
    \includegraphics[width=\textwidth]{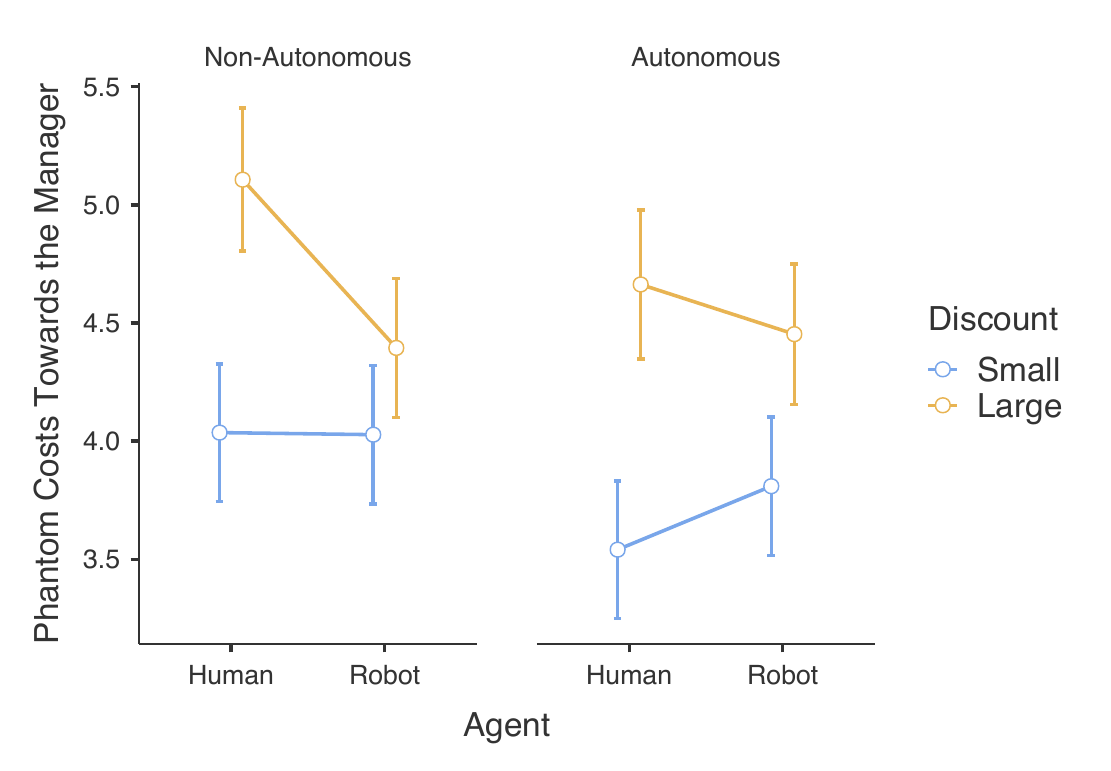}
    \caption{Phantom Costs Towards the Manager}
    \label{fig:Phantom_manager_ANOVA}
  \end{subfigure}
  
  \caption{Estimated Marginal Means of Phantom Costs Perception as a function of Agent, Discount, and Autonomy.}
  \label{fig:Phantom_costs_graphs}
\end{figure*}

We decided to explore phantom costs associated with the car, the sales agent, and the manager separately to better explore how the magnitude of perceived phantom costs may differ according to source. We conducted an ANOVA on the phantom costs as a function of Agent, Discount, Autonomy, and their interactions (\autoref{fig:Phantom_costs_graphs}). Post-hoc comparisons (df = 847) with Tukey's correction were conducted for each significant interaction effect (see \autoref{tab:pc_all} for simple effects).

\subsubsection{Perceptions of the car}
Results revealed a main effect of Agent $(F(1,847) = 5.63, p = .018, \eta_p^2 = 0.007)$, and Discount $(F(1,847) = 195.75, p < .001, \eta_p^2 = 0.188)$, but not Autonomy ($p = .130$). The Agent $\times$ Discount interaction effect was significant ($F(1,847) = 14.43, p < .001, \eta_p^2 = 0.017$), so was the Discount $\times$ Autonomy interaction effect ($F(1,847)$ = 4.23, $p = .040$, $\eta_p^2 = 0.005$). 
Participants perceived more phantom costs associated with the car when there was a large discount rather than a small discount. This was true across all eight conditions.

\paragraph{\texorpdfstring{Agent $\times$ Discount Interaction Effect}{Agent x Discount Interaction Effect}}
When the discount was small, participants perceived few phantom costs. There was no significant difference between humans and robots, regardless of Autonomy. However, when the discount was large, not only did people imagine more phantom costs, but they were significantly more likely to imagine them for a human agent than for a robot, regardless of whether they were autonomous or not.



\paragraph{\texorpdfstring{Discount $\times$ Autonomy Interaction Effect}{Discount x Autonomy Interaction Effect}}
There was no difference among the small discount conditions. However, there was a marginally significant difference where participants perceived more phantom costs associated with the car for a large discount by an autonomous agent than a non-autonomous agent.

\subsubsection{Perceptions of the seller}
Results showed a main effect of Agent $(F(1, 847) = 45.99, p < .001, \eta_p^2 = 0.052)$, Discount $(F(1, 847) = 78.02, p < .001, \eta_p^2 = 0.084)$, and Autonomy $(F(1, 847) = 28.09, p < .001, \eta_p^2 = 0.032)$. Participants perceived more phantom costs for the human than the robot, when the agent made a large discount vs. small, and when the agent was autonomous rather than non autonomous.
The Agent $\times$ Discount interaction was significant, $F(1,847) = 12.94, p < .001, \eta_p^2 = 0.015$.
Perception of phantom costs did not differ between the agents when a small discount was made. When it was large, participants perceived more phantom costs for the human rather than the robot.

\subsubsection{Perceptions of the manager}
Results revealed a main effect of Discount $(F(1, 847) = 55.97, p < .001, \eta_p^2 = 0.062)$ and Autonomy $(F(1, 847)$ = 6.58 $p = .010$, $\eta_p^2 = 0.008)$, but not Agent ($p = .122$). The Agent $\times$ Discount interaction was significant, $F(1,847) = 7.62, p = .006, \eta_p^2 = 0.009$. The Agent $\times$ Autonomy interaction was marginally significant, $F(1,847)$ = 3.31, $p = .069, \eta_p^2 = 0.004$.

\paragraph{\texorpdfstring{Agent $\times$ Discount Interaction Effect}{Agent x Discount Interaction Effect}}
Perception of phantom costs for the sales agent's manager did not differ between agents when a small discount was made. When the discount was large, participants perceived greater phantom costs, particularly for the human's manager rather than the robot's manager.

\paragraph{\texorpdfstring{Agent $\times$ Autonomy Interaction Effect}{Agent x Autonomy Interaction Effect}}
Participants perceived more phantom costs from the manager of the non-autonomous human compared to the autonomous human and autonomous robot. No significant differences were observed between the other comparisons.

\begin{sidewaystable*}
\centering
\caption{Post Hoc Comparisons for Perception of Phantom Costs towards the Car, Agent, and Manager}
\small
\resizebox{\textwidth}{!}{%
\begin{tabular}{lllllllllll}
\toprule
Source & Interaction & Factor 1 & Factor 2 &   & Factor 1 & Factor 2 & $\Delta_{Mean}$ & SE   & t      & p \\
\midrule
Car & Agent $\times$ Discount & Human & Small & - & Human & Large & -1.528 & 0.123 & -12.458 & $< .001$ \\
 &                         &       &       & - & Robot & Small & -0.123 & 0.119 & -1.027  & \phantom{< .}.734 \\
 &                         &       &       & - & Robot & Large & -0.998 & 0.120 & -8.316  & $< .001$ \\
 &                         & Human & Large & - & Robot & Small & \phantom{-}1.405 & 0.123 & 11.434 & $< .001$ \\
 &                         &       &       & - & Robot & Large & \phantom{-}0.530 & 0.124 & \phantom{-}4.288 & $< .001$ \\
 &                         & Robot & Small & - & Robot & Large & -0.875 & 0.120 & -7.279  & $< .001$ \\ \cmidrule{2-11}
 & Discount $\times$ Autonomy & Small & Non-Autonomous & - & Small & Autonomous & \phantom{-}0.047 & 0.119 & \phantom{-}0.390 & \phantom{< .}.980 \\
 &                            &       &                   & - & Large & Non-Autonomous & -1.025 & 0.121 & -8.500 & $< .001$ \\
 &                            &       &                   & - & Large & Autonomous & -1.332 & 0.122 & -10.886 & $< .001$ \\
 &                            & Small & Autonomous & - & Large & Non-Autonomous & -1.072 & 0.121 & -8.886 & $< .001$ \\
 &                            &       &               & - & Large & Autonomous & -1.378 & 0.122 & -11.267 & $< .001$ \\
 &                            & Large & Non-Autonomous & - & Large & Autonomous & -0.307 & 0.124 & -2.482 & \phantom{< .}.063 \\ \midrule
Agent & Agent $\times$ Discount & Human & Small & - & Human & Large & -1.339 & 0.154 & -8.705 & $< .001$ \\
 &                            &       &       & - & Robot & Small & \phantom{-}0.343 & 0.150 & \phantom{-}2.292 & \phantom{< } .101 \\
 &                            &       &       & - & Robot & Large & -0.221 & 0.150 & -1.468 & \phantom{< } .457 \\
 &                            & Human & Large & - & Robot & Small & \phantom{-}1.681 & 0.154 & 10.912 & $< .001$ \\
 &                            &       &       & - & Robot & Large & \phantom{-}1.118 & 0.155 & \phantom{-}7.213 & $< .001$ \\
 &                            & Robot & Small & - & Robot & Large & -0.564 & 0.151 & -3.738 & \phantom{< } .001 \\ \midrule
Manager & Agent $\times$ Discount & Human & Small & - & Human & Large & -1.097 & 0.153 & -7.172 & $< .001$ \\
 &                            &       &       & - & Robot & Small & -0.130 & 0.149 & -0.873 & \phantom{< } .819 \\
 &                            &       &       & - & Robot & Large & -0.635 & 0.150 & -4.247 & $< .001$ \\
 &                            & Human & Large & - & Robot & Small & \phantom{-}0.967 & 0.153 & \phantom{-}6.309 & $< .001$ \\
 &                            &       &       & - & Robot & Large & \phantom{-}0.461 & 0.154 & \phantom{-}2.994 & \phantom{< } .015 \\
 &                            & Robot & Small & - & Robot & Large & -0.505 & 0.150 & -3.371 & \phantom{< } .004 \\ \cmidrule{2-11}
 & Agent $\times$ Autonomy & Human & Non-Autonomous & - & Human & Autonomous     & \phantom{-}0.470 & 0.153 & \phantom{-}3.071 & \phantom{< } .012 \\
 &                            &       &                & - & Robot & Non-Autonomous & \phantom{-}0.361 & 0.150 & \phantom{-}2.398 & \phantom{< } .078 \\
 &                            &       &                & - & Robot & Autonomous     & \phantom{-}0.440 & 0.151 & \phantom{-}2.919 & \phantom{< } .019 \\
 &                            & Human & Autonomous     & - & Robot & Non-Autonomous & -0.109 & 0.152 & -0.717 & \phantom{< } .890 \\
 &                            &       &                & - & Robot & Autonomous     & -0.029 & 0.153 & -0.191 & \phantom{< } .998 \\
 &                            & Robot & Non-Autonomous & - & Robot & Autonomous     & \phantom{-}0.080 & 0.150 & \phantom{-}0.533 & \phantom{< } .951 \\
\bottomrule
\end{tabular}%
}
\footnotesize{\textit{Notes.} Comparisons are based on estimated marginal means. $\Delta_{Mean}$ represents the mean differences.}
\label{tab:pc_all}
\end{sidewaystable*}

\subsubsection{Mediation of Choice through Phantom Costs}
Because the phantom cost survey  items were highly consistent ($\alpha = .860$) and because we did not aim to explore differences among the the items, we used the combined average variable for the mediation analyses. The objective was to determine whether perception of phantom costs mediated participant decision making to buy the car. Autonomy and Discount were the predictors, and we conducted separate analyses for the human and robot agents. Results revealed phantom costs to mediate the relationship between Discount and participant likelihood to buy the car in HHI
$(b = -0.22, SE = 0.03, 95\% CI = [-0.28, -0.16], \beta = -0.23, p < .001)$ and HRI $(b = -0.10, SE = 0.02, 95\% CI = [-0.13, -0.06], \beta = -0.10, p < .001)$.
Large discounts increased the perception of phantom costs, which decreased participant likelihood to buy the car via an indirect effect, although the direct effect of discount remained that it increased the likelihood of purchasing. Perception of phantom costs did not mediate the relationship between Autonomy and Choice for both agents ($p = .408$ in HHI and $p = .142$ in HRI).

\subsection{Likelihood of buying the car}
\begin{figure}
    \centering
    \includegraphics[width=\linewidth]{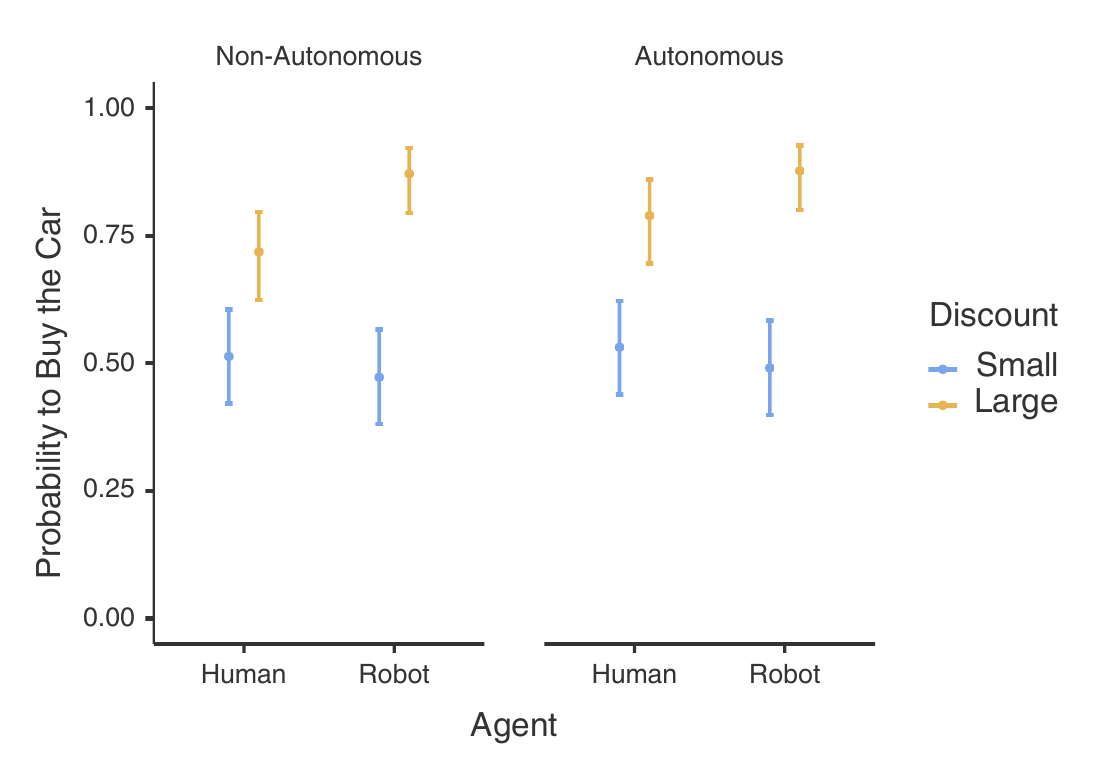}
    \begin{minipage}{0.6\linewidth}
        \centering
        \footnotesize{\textit{Notes.} Error bars correspond to 95\% Confidence Intervals.}
    \end{minipage}
    \caption{Estimated Marginal Means of Buying the Car as a function of Agent, Discount, and Autonomy.}
    \label{fig:Choice_glm}
\end{figure}

We conducted binomial logistic regression analyses to explore the effect of the independent variables (Agent, Discount, Autonomy) and their interactions on participant decisions to buy the car (\autoref{fig:Choice_glm}). Despite elevated perceptions of phantom costs, the odds of accepting an offer were 2.42 times higher when the agent proposed a large vs. small discount $(95\% CI [1.38, 4.30], p = .002)$. Additionally, the Agent $\times$ Discount interaction effect was significant $(\beta = 1.14, p = .011)$. Results were averaged over the levels of Autonomy, with p-value adjustment using Tukey's method to compare a family of four estimates.
Results revealed the odds of buying the car to be 2.82 times higher for a human making a large vs. small discount $(p < .001)$.
Similarly, the odds of buying the car sold by a human with a large discount were 3.32 times higher than the car sold by a robot with a small discount $(p < .001)$.
The odds of buying the car sold by a robot with a large discount were 7.52 times higher than with a small discount $(p < .001)$. The odds of buying a large discounted car sold by a robot were 6.37 times higher than when a small discount was made by a human $(p < .001)$.
The odds of buying a large discounted car sold by a robot were 2.25 times higher than when sold by a human $(p = .012)$.
However, there was no significant difference between the agents making a small discount $(p = .828)$. These findings further support the discount manipulation as part of the study.

\subsection{Self-interest}
A three-way ANOVA and post-hoc analyses were conducted to investigate differences in perceived agent self-interest as influenced by the independent variables (Agent, Discount, Autonomy) and their interactions.
Results revealed a main effect of the Agent $(F(1,847) = 622.50, p < .001, \eta_p^2 = 0.424)$ indicating that the human sales agent $(M = 5.28, SE = 0.07)$ was perceived as more self-interested than the robot sales agent ($M = 2.77, SE = 0.07)$.
Additionally, there was a main effect of Discount $(F(1,847) = 3.87, p = .049, \eta_p^2 = 0.005)$ indicating that agents making a small discount $(M = 4.12, SE = 0.07)$ were perceived as more self-interested that those making a large discount $(M = 3.92, SE = 0.07)$.
Finally, results showed a main effect of Autonomy $(F(1,847) = 86.09, p < .001, \eta_p^2 = 0.092)$ indicating that autonomous agents $(M = 4.49, SE = 0.07)$ were perceived as more self-interested than non autonomous agents $(M = 3.56, SE = 0.07)$.

\subsubsection{Mediation of Phantom Costs Through Self-Interest}
\begin{figure}
    \centering
    \includegraphics[width=\linewidth]{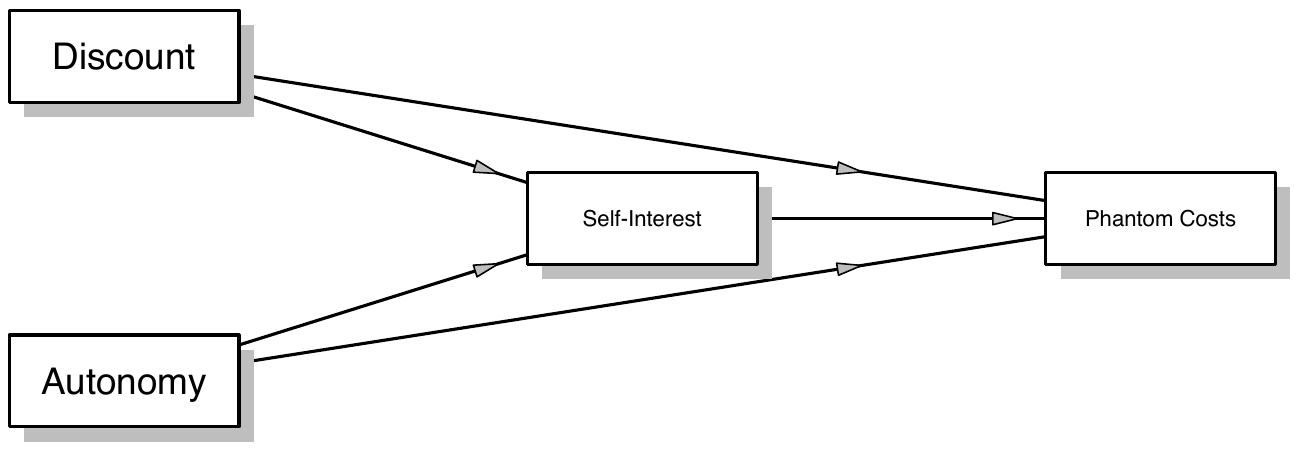}
    \caption{Diagram Paths of the Mediation of Discount and Autonomy on Phantom Costs Perception Through Self-Interest.}
    \label{fig:Mediation_paths_selfinterest}
\end{figure}

Additional analyses were conducted to examine the mediation of phantom costs through self-interest (see diagram paths in \autoref{fig:Mediation_paths_selfinterest}). Using a Generalized Linear Model, we examined indirect, direct, and total effects of Discount and Autonomy as predictors, self-interest as a mediator, and phantom costs as the outcome. Analyses were conducted separately for both types of agents.

\paragraph{Human-Human Interaction}
Analyses for the human agent (N = 420) revealed a significant indirect effect of Autonomy on phantom costs through self-interest $(b = 0.215, SE = 0.050, 95\% CI = [0.117, 0.313], \beta = 0.079, z = 4.311, p < .001)$: autonomy increased perceptions of self-interest $(b = 0.995, SE = 0.131$, $95\% CI$ [0.738, 1.252], $\beta = 0.345, z = 7.587, p < .001)$, which in turn led to greater perceived phantom costs $(b = 0.216, SE = 0.041, 95\% CI [0.135, 0.297], \beta = 0.229, z = 5.239, p < .001)$. The direct effect of Autonomy on the perception of phantom costs was not significant $(b = -0.120$, SE = 0.118, 95\% CI [-0.352, 0.112], $\beta = -0.044, z = -1.014, p = .311)$. The total effect of Autonomy on the perception of phantom costs  was also insignificant $(b = 0.095, SE = 0.115, 95\% CI [-0.130, 0.320], \beta = 0.035, z = 0.830, p = .406)$.

The indirect effect of Discount on perceived phantom costs through self-interest was significant $(b = -0.073, SE = 0.032, 95\% CI = [-0.136, -0.011], \beta = -0.027, z = -2.318, p = .020)$: making a large discount decreased perceived self-interest $(b = -0.339, SE = 0.131, 95\% CI [-0.597, -0.082], \beta = -0.117, z = -2.585, p = .010)$, which in turn led to a lower perception of phantom costs.
The direct effect of Discount on perceived phantom costs was significant $(b = 1.473, SE = 0.112, 95\% CI [1.254, 1.693], \beta = 0.538, z = 13.147, p < .001)$, and so was the total effect $(b = 1.400, SE = 0.115$, $95\% CI [1.175, 1.625]$, $\beta = 0.511, z = 12.184, p < .001)$.

\paragraph{Human-Robot Interaction}
Analyses for the robot agent (N = 435) showed a significant indirect effect of Autonomy on perceived phantom costs through self-interest $(b = 0.066, SE = 0.033, 95\% CI = [0.001, 0.131], \beta = 0.027, z = 1.992, p = .046)$: autonomy increased perceptions of self-interest $(b = 0.879$, SE = 0.151, 95\% CI [0.584, 1.174], $\beta = 0.269, z = 5.836, p < .001)$, thereby increasing the perception of phantom costs $(b = 0.075$, SE = 0.035, 95\% CI [0.006, 0.145], $\beta = 0.100, z = 2.119, p = .034)$. The direct effect of Autonomy on perceived phantom costs was not significant $(b = 0.101, SE = 0.116, 95\% CI [-0.125, 0.328], \beta = 0.041, z = 0.876, p = .381)$. Its total effect was also insignificant  $(b = 0.167, SE = 0.112, 95\% CI [-0.052, 0.387], \beta = 0.068, z = 1.494, p = .135)$.

The indirect effect of Discount on perceived phantom costs through self-interest was not significant $(b = -0.004$, SE = 0.011, 95\% CI = [-0.027, 0.018], $\beta = -0.002, z = -0.356, p = .722)$: making a large discount did not influence the perception of robot self-interest $(b = -0.054, SE = 0.151, 95\% CI [-0.350, 0.241], \beta = -0.017, z = -0.361, p = .718)$. The direct effect of Discount on perceived phantom costs perception was significant $(b = 0.742, SE = 0.111, 95\% CI [0.524, 0.961], \beta = 0.302, z = 6.663, p < .001)$ indicating that the larger the discount, the greater the perception of phantom costs. The total effect of Discount on perceived phantom costs was also significant $(b = 0.738, SE = 0.112, 95\% CI [0.518, 0.958], \beta = 0.301, z = 6.586, p < .001)$.

\section{Discussion}
This study explored the role of agent type, autonomy and self-interest as factors affecting perceptions of phantom costs in both human-human (HHI) and human-robot interactions (HRI). We replicated the phantom costs effect, where people perceived greater risks and ulterior motives for agents making an unreasonably generous offer. This effect was more pronounced when the agent was a human---who was perceived as more self-interested---than a robot. However, despite these perceptions, no monetary backfire effect was observed. Instead, higher discounts increased purchase intentions, consistent with phantom costs increasing more slowly than price appeal as discounts increased; i.e., the net psychological value of a highly discounted car was high despite phantom costs. Moreover, although there was no difference between both agents when making a small discount, participants were more likely to buy the highly discounted car from a robot than a human.
These findings provide a more nuanced understanding of the perception of phantom costs in HHI and HRI, emphasizing the roles of agent autonomy, perceived self-interest, and trust.

\subsection{Self-Interest (H1)}
As predicted, humans were perceived as more self-interested than robots. Although humans and robots were described in vignettes using the same terminology, self-interest may be an attribute of human nature \citep{Haslam2006, Loughnan2007}. Because human nature is what differentiate us from machines \citep{Haslam2006}, participants may rely on this ontological distinction when interacting with a robot. Additionally, agents making a small discount were perceived as more self-interested than those making a large discount. A 5\% discount reduced the actual vehicle price by merely \$100 below the NADA/Blue Book value. This offer may have seemed selfish and  manipulative. Finally, autonomous agents were perceived as more self-interested than non-autonomous ones.

\subsection{Perception of Autonomy (H2)}
Our manipulation of autonomy using the three-facet framework \citep{Kaber2017} was successful, as participants correctly distinguished between autonomous and non-autonomous agents. Despite similar vignette descriptions across both agents conditions, humans were still perceived as more autonomous than robots. This may be because robots are expected to assist humans in general \citep{Xu2024}, or simply that participants used common sense: robots are (presently) less autonomous than humans (see \citep{Gray2007}), even if both were similarly described as autonomous sellers.

Additionally, self-interest partially explained why autonomous agents are perceived as more autonomous (RQ1). Agents described as autonomous were perceived as more self-interested, thereby increasing perceived autonomy. However, the direct effect suggests that other factors influence perceived autonomy. One possibility is that autonomy signals agency to the observer. In particular, the link between self-governance \citep{Kaber2017} and self-control (part of agency \citep{Gray2007}), may influence the perception of self-interest. This suggests that perceived autonomy is not merely functional but also includes moral and social-cognitive assessments of the situation.

\subsection{Phantom Costs}
\subsubsection{Sources of Phantom Costs (H3, H4, H5)}
The three hypotheses were supported. Participants attributed phantom costs to different sources depending on the agent type, discount size, and autonomy.
When the discount was larger, participants were more likely to think something was wrong with the vehicle, especially when sold by a human. However, the agent’s autonomy did not influence the perceived risk of buying the car, as this characteristic was extrinsic to the car. 

Additionally, participants perceived greater phantom costs for the human agent than the robot, when a large discount was made, and when the agent was autonomous rather than non-autonomous. The latter aligns with previous research showing a positive relationship between perceived autonomy and perceived risks \citep{Dinet2022}.
This is because the agent has no reason to be blamed for something it does not control \citep{Gray2007, Furlough2021, Kim2006, Lagnado2008, Monroe2017, Mu2024}. Hence, phantom costs were attributed to the actual decision-maker: the manager.
Interestingly, in any of the hierarchical structures, both the sales agent and the manager were perceived as having ulterior motives. A possible explanation is that they reflect the same entity (i.e., the dealership). Also, non-autonomous robots may appear as a mere extension of the human controlling the robot's behaviors, although they are still attributed some degree of intentionality and risk.

\subsubsection{Mediation of Phantom Costs Through Self-Interest (H6)}
The findings regarding phantom costs were consistent with our hypothesis and with the perception of self-interest. Self-interest mediated the relationship between the discount size and perceived phantom costs for both agent types. However, it only mediated the relationship between the agent’s autonomy and the phantom costs for the human agent, partially supporting our hypothesis.
While the HOSE model predicts that people will perceive more phantom costs from agents who do not follow the norms of self-interest \citep{Vonasch2024}, we found that self-interest increased the perception of phantom costs. However, large discounts still elicited more phantom costs than small discounts, suggesting that phantom costs may relate more on the offer itself than the agent.

\subsubsection{Relationship Between Trust and Phantom Costs (exploratory H9)}
While controlling for each interaction between Agent, Discount, and Autonomy, the perception of phantom costs explained about 50\% of the variance of trust in the agent. The more people trust the agent, the less reason they have to perceive risks and ulterior motives. This is consistent with \cite{Lee2014}, who showed that large discounts elicit greater perceived phantom costs and decrease trust towards the seller, thereby reducing purchase intentions. However, there remains 50\% of the response variance unexplained by the variables in our study (see \citep{Hancock2011} for a list of factors influencing trust).

\subsubsection{Mediation of the Decision to Buy the Car Through Phantom Costs (H7)}
Consistent with the HOSE model \citep{Vonasch2024}, the perception of phantom costs mediated the relationship between the discount and participant decision making. The larger the discount, the greater the perceived phantom costs, and the less likely people were to buy the car. The lack of both a direct and mediated effect of Autonomy through phantom costs highlights the complexity of human decision-making.
The perception of phantom costs and its magnitude depend on the perceived sources of these costs. Although the items assessing phantom costs showed good internal consistency, the variability of phantom costs may mitigate its statistical influence on decision-making.
Autonomy may influence how participants perceived the agent's capacity to make a discount to sell the car, focusing on the \textit{how} question: ``How does the agent make a (generous) discount?''

\subsubsection{Decision-making (H8)}
We examined whether participant decision making was influenced by the type of agent, its level of autonomy, and the discount size.
%
When there was a small discount, there was no difference in participant decision making with the two agents. However, when the car was sold with a large discount, participants  were more likely to buy it from a robot than from a human. This aligns with previous research which showed that people trust robot agents more, increasing people's likelihood to accept their offer \citep{Lebrun2024arXiv, Lebrun2025Replication}. 
Participants were more likely to buy the largely discounted car rather than the one with a small discount. Although the money backfire effect was found in previous research \citep{Vonasch2024, Lebrun2024arXiv}, \citep{Lebrun2024arXiv} did not replicate the effect. This finding, however, still aligns with the HOSE model \citep{Vonasch2024}. Generous offers elicit greater perceived phantom costs. This reduces a person's likelihood to accept the offer, but simultaneously generates greater greed, increasing the likelihood of offer acceptance. The lure of gain can outweigh the perception of risks, encouraging people to accept the offer \citep{Lebrun2024arXiv, Lebrun2025Replication, Vonasch2024}. This finding is consistent with Prospect Theory, which suggests people prefer to optimize gains and avoid losses \citep{Kahneman1979}.
Regarding agent autonomy, the likelihood of buying the car did not appear to be dependent on this factor. Consequently, our hypothesis (H8c) was not supported.  We proffer two reasons for this finding. First, autonomy may not be perceived as a factor influencing decision-making. Second, phantom costs vary in their magnitudes and sources. Large discounts induce greater perceived phantom costs, potentially overshadowing other variables, such as an agent’s autonomy. Given our phantom cost findings, the second option is more likely to explain the current research.

Although previous research showed that a person's income level influences their decision to buy a car \citep{Manski1980, Prieto2013}, we did not find evidence for this in the present study (RQ2). This is likely due to the fact that the participants were not actually buying a car. It does, however, appear, that decision making is influenced by emotional and psychological factors, such as the perception of phantom costs, even in simulated buying scenarios \citep{Prasetyo2020, Vonasch2024}.

\subsection{Limitations}
This study has some limitations.
First, we used ad hoc measures fitting the particular vignettes rather than a statistically validated questionnaire assessing the perception of phantom costs. Although the items showed strong internal consistency, they have not been previously validated. Future research could create a validated questionnaire or use established ones such as the Perceived Danger Scale \citep{Molan2025}.
Second, we did not explore the direct effect of the sufficiency of information on phantom costs and decision-making. Reasonable offers were explained in our vignettes, but the unreasonably generous offers were never justified. Future studies could manipulate justifications for why the human and robot are offering large discounts.
Third, the context of a car dealership may have have affected participants' assessments of the situation. Sales agents might be trusted because dealerships usually provide after-sales support and warranties that could mitigate any potential harm from phantom costs. This could also explain why the monetary backfire effect did not occur. We recommend exploring perceived phantom costs in different contexts in which robots are already used, such as education systems \citep{Belpaeme2018, Lampropoulos2025, Voysey2024} and healthcare \citep{Jayaraman2024, Kyrarini2021, Morgan2022}.
Fourth, conducting the study online limited our findings to our dependent variables, probably failing to capture other types of participant responses and behaviors, as evidenced in \cite{Lebrun2024arXiv}.
Finally, although we used the autonomy framework of \cite{Kaber2017} which defines autonomy through self-governance, independence, and viability, and has gained substantial attention in the human factors and HRI research communities, other frameworks propose different definitions \cite{Kim2024}. Using another framework might provide different conclusions, but it  may not be as effective for clearly instantiating perceptions of autonomous or non-autonomous agents.


\subsection{Implications}
The broader implications of this study include businesses' use of findings as a marketing strategy to enhance sales. By using robots instead of humans to represent brands, businesses may increase customer trust and minimize perceived phantom costs in situations where they wish to offer unusually large discounts.
Similarly, in customer service, chatbots and AI systems could provide sufficient and transparent explanations for their decisions. This could promote effective communication and increase trust in information provided. Phantom costs in this context could include doubts about a chatbot's capability to truly address customer needs, or suspicions about a company's use of a chatbot to deflect responsibilities.
It is also essential to acknowledge the potential for robot owners and operators to misuse this technology to mislead users or, in a more positive way, to obtain efficient HRI that will ensure user safety.

Researchers and designers should ensure that robot behaviors are sufficiently explained, either by the robot itself or another entity. This may help users to accurately perceive  risks and intentions of the agent with whom they are interacting. However, people may not expect the same quantity of information from a robot to justify actions and may doubt less the agent's intentions. Designers may need to provide a way for humans to ask robots to explain their actions.
Additionally, designers and programmers must maintain ethical standards when developing robots. If a designer was to (un)intentionally provide false information regarding agent autonomy and self-interest of a robot, users may incorrectly perceive phantom costs, leading to uncalibrated levels of trust. Such uncalibrated trust may result in robot misuse and a reduction of long-term use. Hence, transparency mechanisms should be incorporated into the robot, clearly stating its capacities and limitations \citep{Angelov2021, Phillips2020}.

\section{Conclusion}
This study tested how agents' self-interest and autonomy influence the perception of phantom costs in HHI and HRI in the context of car sales at a fictitious dealership.
Our results supported the HOSE model, but added nuances to our understanding of how it applies to human-robot interaction. People perceived greater phantom costs when they were offered large discounts on cars, regardless of who the seller was. However, they were somewhat less suspicious of heavily discounted cars when the seller was a robot.
People were particularly skeptical about the motives of whoever was responsible for setting prices and making car sales decisions—i.e., the autonomous agent. When the sales agent was autonomous, people were suspicious of them. However, when the seller was simply implementing a manager's decision rules, participants were more suspicious of the manager, imagining that the manager was responsible for the phantom costs. Thus, people do not necessarily attribute phantom costs to the agent who offers an unreasonable deal. People look beyond the superficial, to glimpse the mind of the agent responsible for the unreasonably generous offer.

A major difference between this current research and previous studies exploring phantom costs was the absence of a monetary backfire effect.
Prior studies found this effect when unreasonably generous offers, such as low prices, led to perceived phantom costs that outweighed a participant's greed for a good deal. The present work did not find any monetary backfire effects. There are many potential reasons for this with the mostly likely being that buying a potentially faulty vehicle is perceived to be a less direct pathway to loss of safety and health than accepting a potentially poisoned cookie from a stranger. However, future research should investigate various contexts in which phantom costs may outweigh greed, and vice versa. This would help explain when phantom costs are likely to produce backfire effects, and when people will acknowledge the risk but still opt to accept a great deal.

Future research should expand this work into non-economic scenarios to understand whether phantom costs remain a key variable in HRI and how they shape user perceptions of agents.
In addition, controlling for agent self-interest would offer a clearer understanding of the impact on perceived phantom costs. For example, in a sales context, the agent could be described as either receiving commissions from sales, attempting to achieve a sales quota, or having no target variable benefits.
Future studies should investigate whether self-interest and/or autonomy are significant predictors of phantom costs and decision-making. For instance, exploring how perceptions of self-interest vary between autonomous and non-autonomous agents could help improve the effectiveness of HRI applications.

\section*{Declarations}
\subsection*{Author contributions}
Conceptualization: BL, AV, DK, CB;
Methodology: BL, AV, DK, CB;
Formal analysis and investigation: BL;
Writing - original draft preparation: BL;
Writing - review and editing: AV, DK, CB;
Resources: CB;
Supervision: CB, AV.

\subsection*{Acknowledgments}
We would like to thank Elena Moltchanova for her support on the statistical analysis.

\subsection*{Data availability}
All data (raw and data analysis) will be made available in a repository upon acceptance.

\subsection*{Conflict of interest}
The authors have no relevant financial or non-financial interests to disclose.

\subsection*{Funding}
No funds, grants, or other support was received.

\subsection*{Compliance with Ethical Standards}
This study was approved by the University of Canterbury Human Ethics Committee (HEC 2019/02 Amendment 11). The participants provided their written informed consent to participate in this study.

\appendix

\section{Relationship between phantom costs and trust}
Linear regression analyses were conducted to explore the relationship between Phantom Costs perception and Trust for each interaction between Agent (Human vs Robot) and Discount (Small vs Large). Trust was the outcome and Phantom Costs, Agent, Discount, and Autonomy were the predictors. Main effects, two-way and three-way interactions were included in the model. The model was significant, $F(15,839) = 58.22, p < .001$. The adjusted R square indicated that the predictors of this model explained 50.12\% of the variance of the dependent variable Trust.
Results revealed that phantom costs perception significantly predicted how trustworthy the agent was perceived to be. This means that the more phantom costs participants perceived, the less they trusted the agent $(\beta = -0.53, p < .001)$. The other predictors were not significant predictors of trust $(p > .05)$.

\section{RQ1: Exploratory Mediation Analysis Through Self-Interest}
Exploratory mediation analyses were conducted to examine whether the perception of self-interest mediates the relationship between Autonomy and Perceived Autonomy. Results revealed an indirect effect of Autonomy on Perceived Autonomy through self-interest was significant $(b = 0.260, SE = 0.042, 95\% CI = [0.179, 0.342], \beta = 0.088, z = 6.27, p < .001)$, indicating that autonomous agents led to an increase of self-interest, which in turn led to a higher perceived autonomy. This is explained by looking at the components. The effect of Autonomy on Self-Interest was significant $(b = 0.924, SE = 0.132, 95\% CI = [0.665, 1.184], \beta = 0.232, z = 6.98, p < .001)$. Additionally, the effect of Self-Interest on Perceived Autonomy was significant $(b = 0.282, SE = 0.020, 95\% CI = [0.243, 0.320], \beta = 0.378, z = 14.30, p < .001)$. The direct effect of Autonomy on Perceived Autonomy was also significant $(b = 1.370, SE = 0.078, 95\% CI = [1.216, 1.523], \beta = 0.461, z = 17.48, p < .001)$. The total effect (including indirect and direct effect) was also significant $(b = 1.630, SE = 0.085, 95\% CI = [1.464, 1.527973], \beta = 0.594, z = 19.20, p < .001)$.

\section{Effect of Income on Decision-Making}
We included the main effect of Income in the model above to explore whether participants' income of the last 12 months influenced their decisions to buy the car. There was no significant main effect of any of the Incomes, with a p-value ranging from .166 to .993.

\bibliographystyle{plain}
\bibliography{manuscript_sample}

\end{document}